\documentclass[journal]{IEEEtran}

\usepackage{cite}
\usepackage{blindtext}
\usepackage{amsmath,amssymb,amsfonts}
\usepackage{graphicx}
\usepackage{textcomp}
\usepackage{array}

\usepackage{ragged2e}
\usepackage{amsthm}
\usepackage{array}
\usepackage{tabulary}
\usepackage{balance}
\usepackage{soul,color}
\usepackage{algorithm}
\usepackage[noend]{algpseudocode}
\usepackage[utf8]{inputenc}
\usepackage[export]{adjustbox}
\usepackage[T1]{fontenc}
\usepackage[english]{babel}
\usepackage[utf8]{inputenc}
\usepackage{graphicx}
\usepackage{amssymb}
\usepackage{amsmath}
\usepackage{mathtools}
\usepackage{amsfonts}
\usepackage{array}
\usepackage{cite}
\usepackage{amsbsy}
\usepackage{amsmath}
\usepackage{amsfonts}
\usepackage{amssymb}
\usepackage{latexsym}
\usepackage{epsfig}
\usepackage{pdfpages}
\usepackage{color}
\usepackage{soul}
\usepackage{caption}
\usepackage{subfigure}
\usepackage{setspace}
\usepackage{multirow}
\usepackage{array}
\usepackage{commath}
\usepackage{diagbox}
\usepackage{textcomp}
\usepackage[table,xcdraw]{xcolor}
\def\BibTeX{{\rm B\kern-.05em{\sc i\kern-.025em b}\kern-.08em
		T\kern-.1667em\lower.7ex\hbox{E}\kern-.125emX}}
\ifCLASSINFOpdf
\else
\fi
\begin{document}
%
	\title{Identification of Smart Jammers: Learning based Approaches Using Wavelet Representation}
\author{Ozan~Alp~Topal, Selen~Gecgel, Ender~Mete~Eksioglu \IEEEmembership{Member, IEEE}, and Gunes~Karabulut~Kurt, \IEEEmembership{Senior Member, IEEE}
\thanks{The authors are with Department	of Electronics and Communication Engineering, Istanbul Technical University, 34469 Istanbul, Turkey (e-mail: topalo@itu.edu.tr; gecgel16@itu.edu.tr; eksioglue@itu.edu.tr;gkurt@itu.edu.tr).}
\thanks{This work was supported in part by the  Turkcell İletişim Hizmetleri A.Ş. under Grant BSTB032384.}}

%
%

\markboth{Journal of \LaTeX\ Class Files,~Vol.~14, No.~8, August~2015}%
{Shell \MakeLowercase{\textit{et al.}}: Bare Demo of IEEEtran.cls for IEEE Journals}

\maketitle

\begin{abstract}
	Smart jammer nodes can disrupt communication between a transmitter and a receiver in a wireless network, and they leave traces that are undetectable  to classical jammer identification techniques, hidden in the time-frequency plane. These traces cannot be effectively identified through the use of the classical Fourier transform based time-frequency transformation (TFT) techniques with a fixed resolution. Inspired by the adaptive resolution property provided by the wavelet transforms, in this paper, we propose a jammer identification methodology that includes a pre-processing step to obtain a multi-resolution image, followed by the use of a classifier. Support vector machine (SVM) and deep convolutional neural network (DCNN) architectures are investigated as classifiers to automatically extract the features of the transformed signals and to classify them. Three different jamming attacks are considered, the barrage jamming that targets the complete transmission bandwidth, the synchronization signal jamming attack that targets synchronization signals and the reference signal jamming attack that targets the reference signals in an LTE downlink transmission scenario. The performance of the proposed approach is compared with the classical Fourier transform based TFT techniques, demonstrating the efficacy of the proposed approach in the presence of smart jammers.  
\end{abstract}


%
\IEEEpeerreviewmaketitle

\section{Introduction}
\IEEEPARstart{O}{rthogonal} frequency division multiple access (OFDMA) technique constitutes the physical layer multiplexing method of choice for the Long Term Evolution (LTE) networks and its later {Releases} due to its robustness against noise and fading impacts \cite{OFDMJamming}. Based on the advantages provided by OFDMA, LTE has brought up numerous favorable properties including but not limited to higher data rates, better coverage, higher energy efficiency and lower latency than previous cellular networking technologies \cite{baker2012lte}. Despite the advantages that come with the OFDMA technique, LTE networks suffer from attacks generated by active radio nodes \cite{LTESec}. These attacks generally include cases where an attacker (a jamming node) transmits a signal aiming to falsify the receiver or disrupt the communication. Such communication disruptions are frequently referred to as layer-1 denial-of-service (DoS) attacks.

Jamming attacks have been one of the conspicuous issues in wireless communication networks. As communication systems become an integral part of our daily lives, transmission of the critical information is expected to be more reliable, increasing the effective impact of the jamming attacks. These attacks can be initiated in various ways through different types of jamming nodes. In order to avoid or to mitigate the effects of the jamming, it  is critical to detect the existence of a jamming node and to detect its attack methodology. These processes are respectively named as jammer detection and jammer identification.

As detailed in the following section, jammer identification is usually based on the network measurements or signal characteristics. On determining the attack type, identification algorithms may use different probabilistic approaches. Machine learning techniques also gain importance on identification and detection algorithms. Especially, deep convolutional neural network (DCNNs)  have been studied in  different communication problems that require quick response in  real-time applications. The main application areas can be listed as localization \cite{S13}, modulation classification \cite{S14}, channel decoding \cite{S16} and waveform recognition \cite{zhang2017convolutional}.

A common step in all of these detection and classification methods is a time-frequency transformation (TFT) based pre-processing which is capable of capturing the jamming effects on the received signal. Most commonly used TFT method is spectrogram and has been utilized for different signal analysis purposes \cite{boashash2015time}. Spectrogram is based on short time Fourier transform (STFT) and represents the signal on the time-frequency plane. As it will be detailed below, spectrogram uses fixed window size which results in fixed resolution of the signal on time-frequency plane.

Concurrent with the evolution of communication networks, newly introduced jammer types may attack in  { very short time intervals or narrow and changing frequency bands}. In that way, jamming nodes are able to hide in the time-frequency plane and become invisible to conventional TFT methods such as STFT. These kind of attackers are named as smart jammers, and their effects are already shown in numerical and measurement studies \cite{smart-USRP, zuba2011launching}. Reduced hardware costs with easily accessible software enable smart jammers to observe the transmissions and detect the vulnerable parts of the transmitted packets, possibly targeting the reference signals. With these observations, smart jammers gain an advantage over the legitimate nodes as shown in \cite{Az2017}. Defense mechanisms against jamming attacks are proposed from different perspectives such as  repeated game  algorithms \cite{Az2017}, or frequency hopping algorithms \cite{shridhara2002jamming}. These network strategy algorithms require the knowledge of the jammer type in order to combat against the jammer attack. Therefore, the jammer identification is a requirement prior to the jamming prevention/anti-jamming systems. 

Motivated by this problem, we propose a wavelet tranformation based pre-processing for smart jammer identification in order to improve detection and classification performance. Wavelet transforms are commonly used signal processing techniques thanks to the tunable time and frequency resolution properties \cite{wavelet}. Similar to Fourier transform, wavelet transforms represent a signal through a linear combination on particular basis functions. In Fourier transform, the basis functions are sinusoidals which are not localized on time-frequency plane due to continuous oscillation. On the contrary, wavelets are finite in time and localized on time-frequency plane, enabling successful detection and identification of the smart jammers hiding in the time-frequency plane.

\begin{figure}[tb]
	\centering
	\includegraphics[width=0.35\linewidth]{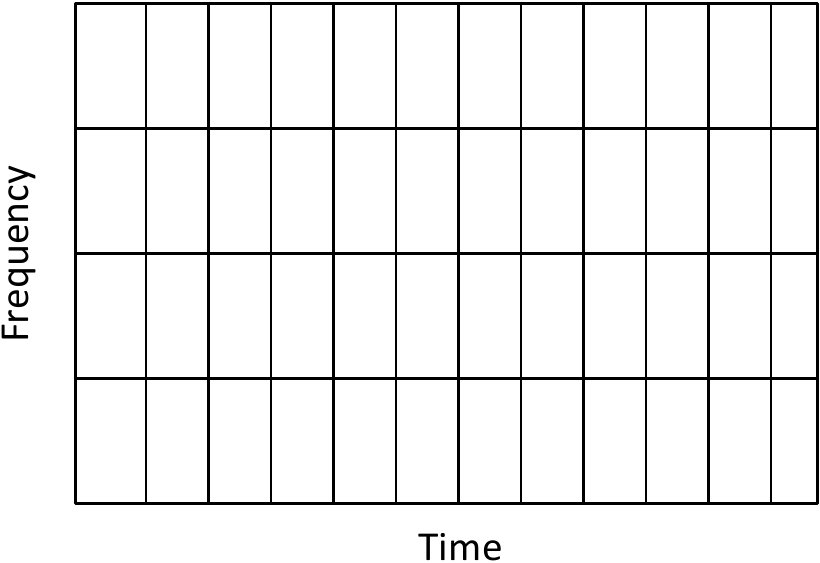} \hspace{1cm}
	\includegraphics[width=0.35\linewidth]{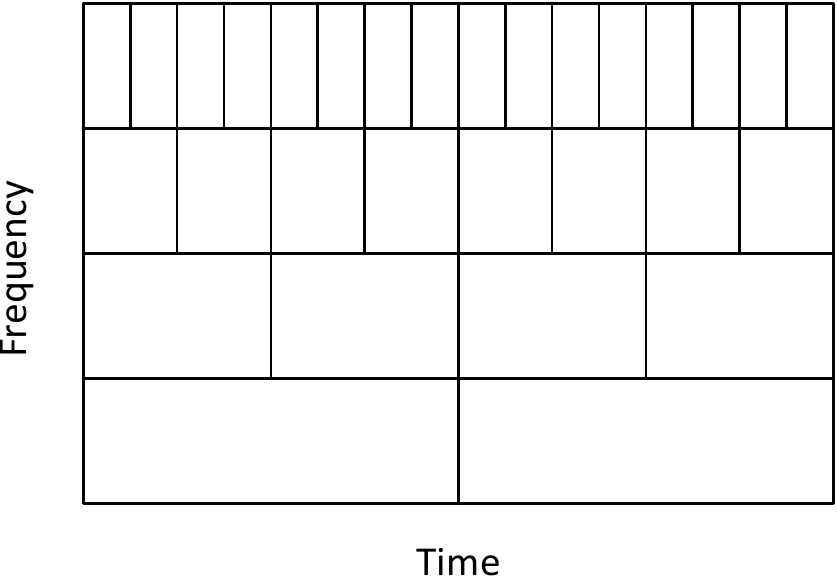}
	
	\caption{An exemplary graphical representation of (a) short time Fourier transform (STFT) and (b) wavelet transform.}
	\label{fig:morelet}
\end{figure}
STFT provides an observation on temporal changes of the signal by applying a rectangular window to the signal before Fourier transform. Since same rectangular window is applied to the signal in STFT, time-frequency resolution of the signal would be same at all positions. Unlike STFT, varying window size in the wavelet transforms provides multi-resolution on the different positions of the signal \cite{strang1993wavelet}, as highlighted in Figure \ref{fig:morelet}. In this way, we can provide an adaptive resolution property against changing signal characteristics resulting from smart jammers.
\begin{figure}[tb]
	\centering
	\includegraphics[width=0.75\linewidth]{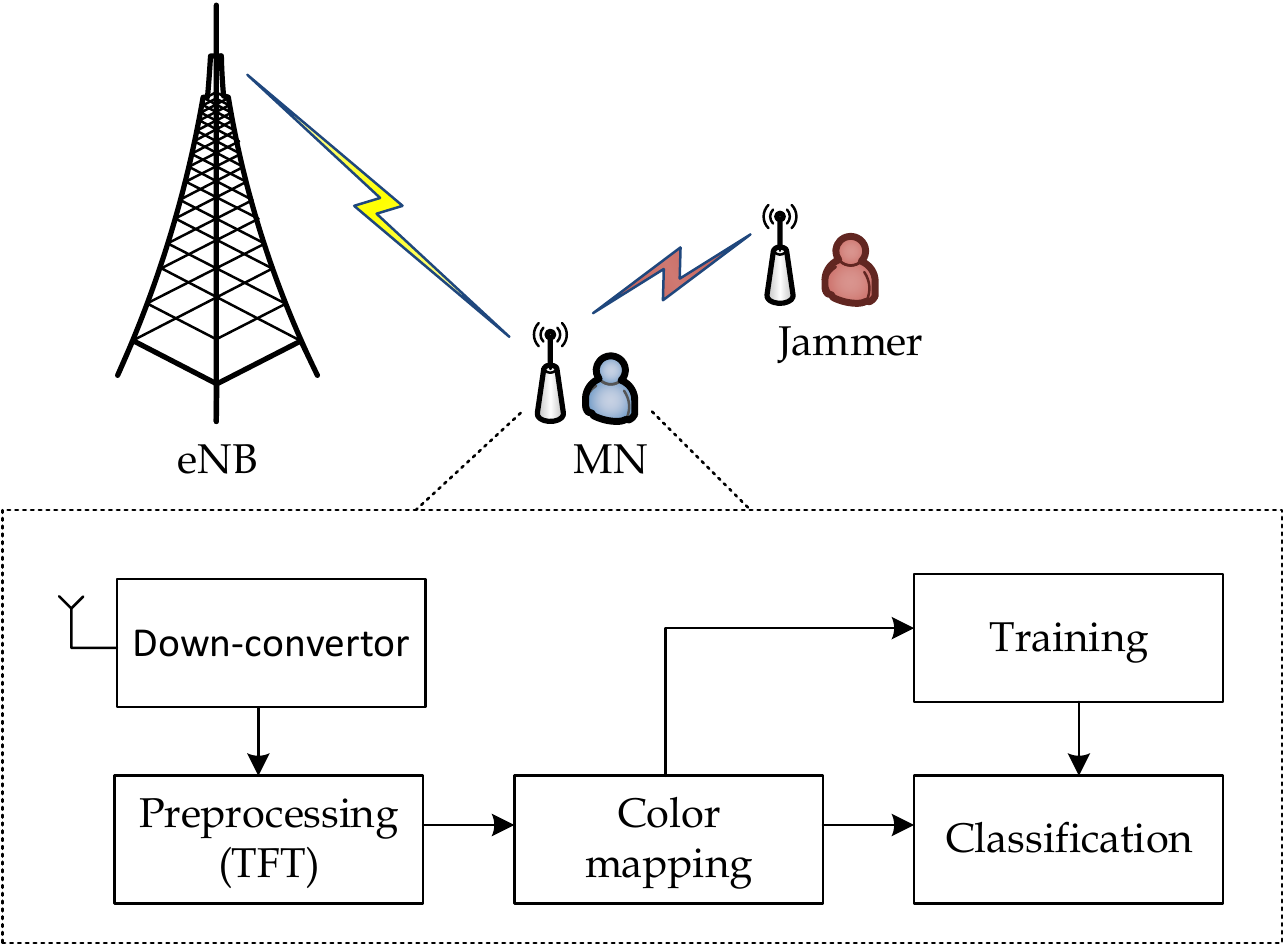}
	\caption{A base station (eNB), monitoring node (MN) and a jammer in a LTE cell. The received signal from the MN would be the superposition of LTE signal and jammer signal.}
	\label{fig:cell}
\end{figure}
\begin{table*}[t]
	\footnotesize
	\caption{Literature review of the anti-jamming techniques.}
	\label{table:literature}
	\begin{tabular}{|c|c|c|c|}
		\hline
		\rule[-1ex]{0pt}{2.5ex}  Research target             &                Reference                 &               Attack Type               &                  Method                   \\ \hline\hline
		\rule[-1ex]{0pt}{2.5ex} \multirow{6}{*}{jamming detection }  &    \cite{firouzbakht2014performance}       &    MAC layer smart jammer attacks                    &             Game theory             \\ \cline{2-4} 
		\rule[-1ex]{0pt}{2.5ex}          &     \cite{Chen2018}                &          MAC layer smart jammer attacks            &           Time series analysis            \\ \cline{2-4}
		\rule[-1ex]{0pt}{2.5ex}            &     \cite{Li2007}           &          smart jammer attacks                 &             Game theory             \\ \cline{2-4}
		\rule[-1ex]{0pt}{2.5ex}          &   \cite{Kurt2018}         &     hybrid jammer attacks in smart grid          &              Kalman filters               \\ \cline{2-4}
		\rule[-1ex]{0pt}{2.5ex}             &     \cite{S8}                            &        continuous single-tone attack             &                    SVM                    \\ \cline{2-4}
		\rule[-1ex]{0pt}{2.5ex}      &       \cite{tugnait2018pilot}          &  pilot spoofing on MISO system &     random matrix theory, beamforming     \\ \hline
		\rule[-1ex]{0pt}{2.5ex}    \multirow{8}{*}{jamming identification }      &         \cite{Az2017}                  &  MAC layer attacks            &             Game theory             \\ \cline{2-4}
		\rule[-1ex]{0pt}{2.5ex}               &    \cite{Kar2018}             &   jamming in vehicle communication   &   unsupervised learning with clustering       \\ \cline{2-4} 
		\rule[-1ex]{0pt}{2.5ex}              &  \cite{S4}                &      MAC layer attacks       &                PSR \& PDR                 \\ 	\cline{2-4}	         
		\rule[-1ex]{0pt}{2.5ex}          &     \cite{Junfe2018}    &           BJ attacks on SAR          &                    CNN                    \\ \cline{2-4}
		\rule[-1ex]{0pt}{2.5ex}             &      \cite{S9}          &  MAC layer attacks         &             Machine learning              \\ \cline{2-4}
		\rule[-1ex]{0pt}{2.5ex}       &               \cite{w2017jamming}             &             MAC layer attacks          &                    CNN                    \\ \cline{2-4}
		\rule[-1ex]{0pt}{2.5ex}          &        \cite{Zhang2012}        &       attacks on time varying channel        &        jamming coherence bandwidth        \\ \cline{2-4} 
		\rule[-1ex]{0pt}{2.5ex}               &   our solution            &  smart jammer attacks at LTE physical layer   &   TFT pre-processing with DCNN classifier       \\ 
		\hline			        				
		\rule[-1ex]{0pt}{2.5ex} \multirow{9}{*}{jammer countermeasure }            &     \cite{halim}                    &     LTE PSS\&SSS attack              &            Adaptive filtering             \\ \cline{2-4}
		\rule[-1ex]{0pt}{2.5ex}  &       \cite{shahriar2013performance}   &     OFDM smart jamming        &            pilot randomization            \\ \cline{2-4}
		\rule[-1ex]{0pt}{2.5ex}           &  \cite{Ak2017}  &  pilot distortion attack in massive MIMO system 
		Detection      &       multiple-antenna CFO estimate       \\ \cline{2-4}
		\rule[-1ex]{0pt}{2.5ex}            &    \cite{Nav2007}       &        jamming              &             frequency hopping             \\ \cline{2-4}
		\rule[-1ex]{0pt}{2.5ex}         &         \cite{Xia2018c}           &                          NOMA jamming       & Game theory, reinforcement learning \\ \cline{2-4}
		\rule[-1ex]{0pt}{2.5ex}         &         \cite{harvesting-mitigation}           &                        SJ attacks on wireless energy harvesting       & Markov decision process \\ \cline{2-4}
	    \rule[-1ex]{0pt}{2.5ex}           &    \cite{Wan2018}                    &       Cyber-Physical System               &             Stackelberg Game              \\ \cline{2-4}
		\rule[-1ex]{0pt}{2.5ex}     &    \cite{xiao2018uav}       & smart jamming on VANETs    & Game theory, reinforcement learning \\  \cline{2-4}
		\rule[-1ex]{0pt}{2.5ex}           &    \cite{deep-mitigation}                    &       generative adversarial network and mitigation               &             deep learning             \\\hline
	\end{tabular}
\end{table*}
We consider the LTE-downlink channel jamming scenario, where a stationary monitoring node (MN) captures downlink channel observations by continuously processing the received baseband signal. The MN is synchronized with the cell and a stationary jammer attacks the transmission from the base station (eNB), as shown in Figure \ref{fig:cell}. Note that, MN does not have any prior knowledge about the jammer attack. In the presence of a smart jammer, two major vulnerabilities become apparent in LTE-downlink transmission: synchronization signals and reference signals \cite{lichtman2013vulnerability}. These signals are transmitted along with the message signal over the network. Unlike message signals, they contain critical information and even a partial disruption of this information may cause loss of the complete LTE packet. Our main contributions can be listed as the following;
\begin{itemize}
	\item We introduce a novel system model for the identification of the smart jammers. Proposed identification system can be divided
	into two steps: $(i)$ a pre-processing step to highlight the
	disrupted parts of the signal, $(ii)$ a classification step that
	automatically identifies  jamming signals. 
	\item We provide a wavelet based pre-processing step that conveys multi-resolution representation of the signals. The validity of the proposed pre-processing step is confirmed via simulations. We compare Gabor wavelet transform with TFT's that were previously used for classification pre-processing: spectrogram and Choi-Williams.
	\item For the classification step, we propose a DCNN architecture to automatically
	extract the features of the transformed signals and to classify
	them.
	\item  As an alternative for the classification step, we utilized SVM also for the first time considering the smart jammer identification. The classification accuracies of two classification schemes are compared with each other.
	\item The proposed identification scheme is repeated for different location cases considering that the identification accuracy could change with respect to distances between the jammer and the MN or between the eNB and the MN.
\end{itemize}

The rest of this paper is organized as follows. In Section II, we discuss jammer detection, classification and mitigation literature. In Section III-A, we briefly review some relevant features of LTE downlink signal model. In Section III-B, we discuss possible jammer attack types. In Section IV, we present the proposed system model for jammer identification. Numerical analysis and simulations are presented in Section V, and the open issues are given in Section VI. Finally, the paper is concluded in VII.

\section{Related Work}
As presented in Table \ref{table:literature}, anti-jamming literature can be reviewed into several different research focuses. In \cite{firouzbakht2014performance}, the authors consider reactive jamming attacks on the packet based transmission network. They present an anti-jamming strategy based on zero-sum game model at the media access control (MAC) layer. The authors of \cite{Chen2018} present a smart jamming detection algorithm based on the time-series analysis. In \cite{Li2007}, the authors propose a cross-layer jammer detection and prevention system that utilizes game theoretical model. \cite{Kurt2018} presents a real time smart jamming detection system by utilizing Kalman filters. [19] shows that by using SVM classifier the performance of jamming detection improves and hardware complexity decreases. The authors of [20] consider pilot jamming and spoofing attacks, and they propose a random matrix  theory-based  source enumeration approach for attack detection. 

Besides jammer detection, identification of a specific attack type is also a critical part of the anti-jamming strategies. In [11], the authors propose a jammer identification system that differentiates the bandwidth characteristics of different jamming types. In \cite{Kar2018}, the authors propose unsupervised learning algorithms that  identify different type of jamming attacks on vehicle communication systems.  \cite{S9} presents a machine learning-based jamming identification approach for IEEE 802.11 networks, where the authors compare the accuracy and robustness of different popular machine learning techniques, which include decision tree, adaptive boosting, support vector machine (SVM), expectation maximization and random forest. Jammer identification is usually based on network measurements such as packet delivery ratio \cite{S3}, signal strength \cite{S4}, packet transmission ratio and erroneous packet ratio \cite{S5}.

In addition to shallow machine learning based identification schemes, the use of DCNNs have been studied in  different communication problems that require quick response in  real-time applications \cite{deep-main}. In [23], the proposed system model can identify different types of barrage jamming attacks using CNN as a classifier.  The most related paper with our work, \cite{wu2017jamming} proposes a CNN based jammer identification method where jammer types have been classified solely depending on the attacked bandwidth.
In our work, we consider smart jammer attacks at the physical layer of an LTE network. The proposed system model can identify 3 different jamming types as well as inactive jammer case.

Jammer countermeasure is another significant part of the anti-jamming strategies. While some countermeasures do not require the knowledge of the attacker type, some countermeasures require the knowledge of the attacked bandwidth or the attacker strategy. In this case, these systems require a jammer identification process before performing their strategies. In [28], the authors propose an adaptive filtering method considering the reference signal attacks to the LTE networks. Their methodology does not require an identification process. Similarly, in [30], the authors utilize the multiple antenna diversity to combat the jamming attacks without any identification process. However, in [31], the legitimate nodes benefit from the knowledge of the jammer type and existence. In [32, 34], the jammer identification gives an advantage to the legitimate nodes for combating the jamming attacks. The authors of \cite{deep-mitigation} initiate a jamming attack using a deep neural network and propose mitigation methods for this type of attack. As a main difference from other machine learning based anti-jamming systems, we investigate the effect of the different pre-processing methods on the learning performance of the machine learning methods.

\section{Background}
This section is devoted to the review of relevant characteristics of LTE downlink channel model and jamming attack models. As previously mentioned, smart jammers would require some insights about the physical layer to focus on the vulnerabilities of the system. Therefore, before describing jamming attack models, the physical layer properties of an LTE system should be addressed.
\subsection{LTE - PHY Model}
\begin{figure}[tb]
	\centering
	\includegraphics[width=\linewidth]{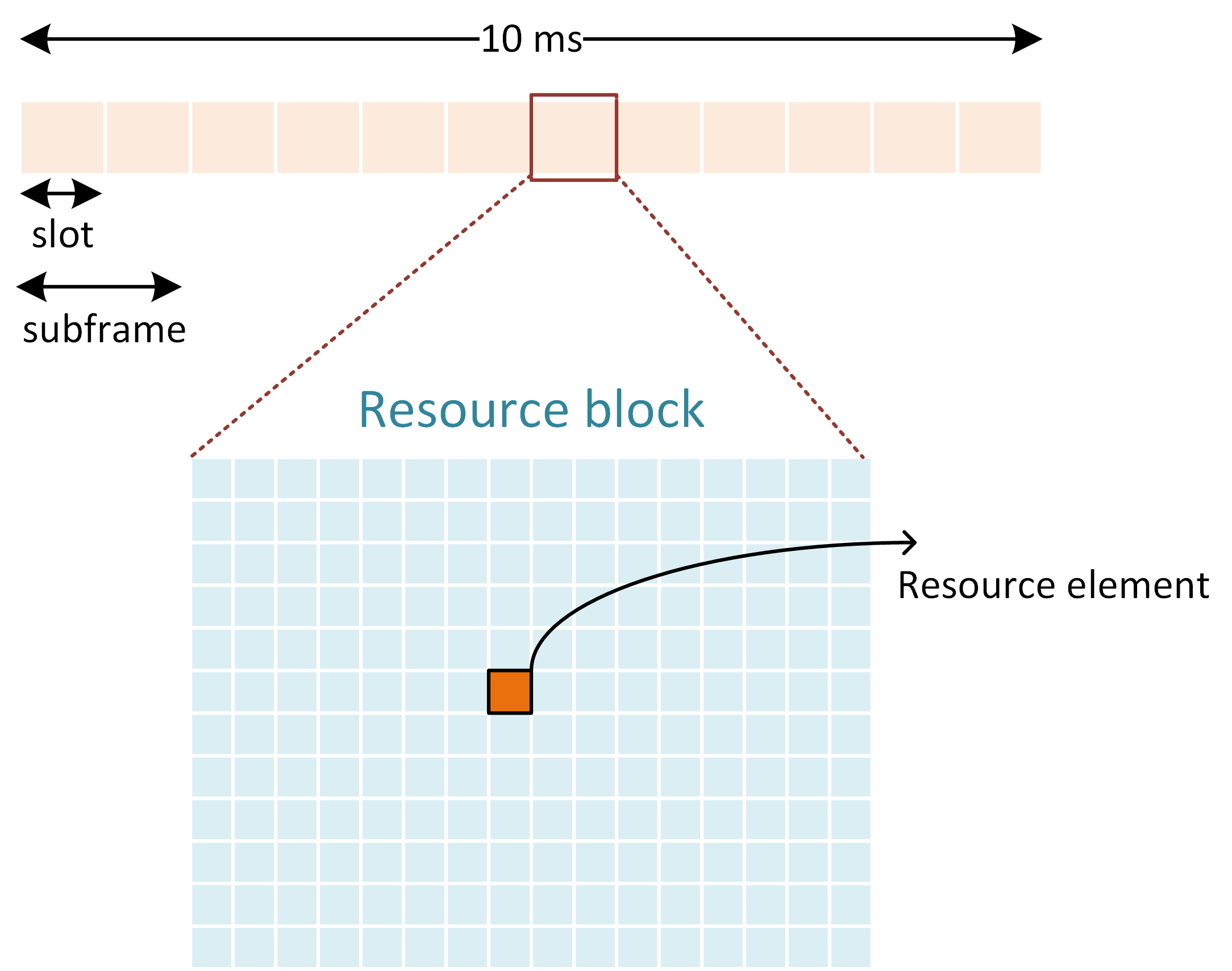}
	\caption{LTE frame, subframes and time slots.}
	\label{fig:LTE-frame}
\end{figure}

LTE downlink channel uses OFDMA as the channel access scheme. {The transmission can be in frequency division duplex (FDD) or time division duplex (TDD) mode \cite{etsi2009136}.} Figure \ref{fig:LTE-frame} shows an illustration of an LTE frame considering FDD transmission mode. Data is delivered in frames of 10 milliseconds. An LTE frame is composed of ten subframes of 1 ms, and each subframe contains 2 time slots of 0.5 ms.
Each time slot contains $L$   symbols, and $L$ depends on the cyclic-prefix (CP) mode. $K$ denotes the total number of subcarriers. Depending on the transmission bandwidth, $K$ changes between 128 to 2048.  The smallest defined unit, $a_{k,l}$, denotes a resource element which consists of $k^{\text{th}}$ subcarrier during the $l^{\text{th}}$ symbol, where $k=0, 1, \cdots, K-1$ and $l=0, 1, \cdots, L-1$. According to the selected bandwidth and protocol, control and message signals are mapped into empty resource blocks. 

From an attacker perspective, obtaining the exact subcarrier and symbol index of the control signals can be beneficial because attacker can damage the transmission by  attacking only the resource elements where control signals are mapped. The remaining part of this section will focus on the generation and mapping of the control signals in LTE downlink channel.

\subsubsection{Synchronization Signals}
In order to determine and to synchronize to a cell, a user equipment (UE) needs to acquire the frame timing information, estimate the carrier frequency offset (CFO) and also identify the cell. For this purposes, eNB generates and maps two signals into resource elements: primary synchronization signal (PSS) and secondary synchronization signal (SSS).

eNB maps the PSS in the last symbol of first time slot of the first subframe (subframe 0). The PSS is also mapped in subframe 5 which means UE can be synchronized on 5 ms basis. SSS symbols are also mapped in the same subframe of PSS with the same subcarrier indexes but a symbol before the PSS.

To generate the PSS and the SSS, the transmitter uses a complex valued Zadoff Chu (ZC) sequence \cite{etsi2009136}. Let $d_u(m)$ denote the ZC sequence of the PSS and the SSS,
\begin{equation}
d_u(m)= \begin{cases}
e^{\frac{-j\pi u m(m+1)}{63}}\text{ , } &  0\leq m \leq 30\\
e^{\frac{-j\pi u (m+1)(m+2)}{63}}\text{ , } & 31\leq m \leq 61
\end{cases},
\end{equation}
where $u$ is the root value for the ZC series and changes with the index of the symbol.
In this case $d_u(m)$ is mapped according to the relevant resource elements;
$$
a_{k_s, l_s}=d_u(m),
$$
where $m=0,1,\dots,61$  and $u\in \{0, 10\}$. $k_s$ denotes the subcarrier indices of mapped resource elements with $k_s=m-31+K/2$. $l_s$ denotes the symbol indexes of the mapped resource elements and set to $L-2$ for the PSS and $L-1$ for the SSS.

At the UE, first PSS is extracted for time frame and frequency synchronization. In the next step UE extracts the SSS. Through the SSS, the UE extracts about the CP mode and the duplexing mode used by the cell. From their combination, the mapped locations of the reference signals can be found in order to realize the channel estimation and equalization steps. In case of any disruption on the PSS and SSS, not only the synchronization at the UE is affected, but also the packet will get corrupted because of the erroneous channel estimation \cite{halim}.
\subsubsection{Reference Signals}
Reference signals are periodically transmitted in LTE networks to perform channel estimation and frequency domain equalization at the receiver side after CP removal and  demodulation steps.
The cell identification determines time and frequency domain locations of the quadrature phase shift keying (QPSK)  modulated reference signals. In this work, subcarrier and symbol indices of the reference signals are assumed as fixed, respectively as $k_{rf}=0,7,14,\cdots,N$ and $l_{rf}=0,7,14,\cdots,N$. Each  $a_{k_{rf},l_{rf}}$ is filled with a random  complex number, $a_{k_{rf},l_{rf}}\in\{\frac{1+j}{\sqrt{2}},\frac{1-j}{\sqrt{2}},\frac{-1+j}{\sqrt{2}},\frac{-1-j}{\sqrt{2}}\}$. UE is assumed to know the index values $k_{rf}$, $l_{rf}$ as well as the values of the reference signals. 
The receiver produces an initial estimate for the channel coefficients of the received signal by  utilizing the reference signals.  Afterwards, these initial estimates are interpolated by using the additional information coming from the other resource elements, and the channel coefficient estimates are finalized. Following the channel estimation, UE recovers the received packet by using the estimated channel coefficients for equalization.  If the estimated coefficients are erroneous, then  coefficients belonging to other resource elements will also be inaccurate after interpolation. Additional control signals may also get mapped into resource blocks \cite{etsi2009136}. Yet, in our model, resource elements are filled by the union of the resource elements filled with synchronization signals, reference signals and message signals.  The mapped subcarrier and symbol indexes for message signal can be denoted  with $k_m$ and $l_m$, respectively.

After mapping all signals into resource blocks, discrete time baseband representation of the transmitted downlink LTE signal $s(l)$ can be obtained as
\begin{equation}
s(l)=\frac{1}{\sqrt{K}}\sum_{k=0}^{K-1}{a_{k,l}e^{j2\pi k l/K} \omega \left( {l}/{K+K_{CP}-1} \right)},
\end{equation}
where $K_{CP}$ denotes the cyclic-prefix length and $\omega(l')$ is a discrete rectangular window that is defined by

\begin{equation}
\omega(l')=
\begin{cases}
1& \text{ ; } 0\leq l'\leq1 \\
0& \text{ ; otherwise}
\end{cases}.
\end{equation}

\subsection{Jamming Attacks}
Jamming attacks can be classified according to different characteristics such as attacked bandwidth and jamming signal transmission. We will focus on jammers that attack the LTE characteristics explained above. Therefore four different types of jamming cases are considered: ($i$) there is no attack to the system, ($ii$) the complete frequency band is attacked, ($iii$) only the synchronization signals are attacked and ($iv$) only reference signals are attacked. Following titles are the specific names for these different jammer types respectively.

\subsubsection{Barrage Jamming (BJ)}
Barrage jamming (BJ) is the most frequent form of the jamming attack. Its detection \cite{detection-1}, \cite{detection-2} and mitigation \cite{mitigate} have been exhaustively discussed in the literature. In BJ, the attacker continuously transmits band limited noise over the entire spectrum of the receiver.  Note that with enough observations over the network, attacker may transmit noise only during the transmission of a specific signal \cite{mietzner2012responsive}. This approach requires more complexity or additional information with respect to the network, but it reduces the energy consumption of the jammer \cite{1}. With additional knowledge on the network, BJ attacks can further be diversified, but in this work BJ is assumed to transmit noise over the entire bandwidth of the receiver. Discrete time representation of the transmitted BJ signal is $$j_b(l)\sim \mathcal{CN}(0,\sigma_j^2),$$ where $\mathcal{CN}(\mu,\sigma^2)$ denotes the complex normal distribution with mean $\mu$ and variance $\sigma^2$.
Even though BJ does not require any prior knowledge of the network, in order to initiate the following attacks, an attacker will require to extract some information about the LTE network.
\subsubsection{Synchronization Signal Jamming (SSJ)}
Unlike BJ, in synchronization signal jamming (SSJ), the attacker should be aware of the locations of the resource blocks as explained above. Although there are several ways of jamming synchronization signals, we assume that the attacker generates an LTE frame similar to the transmitter, but maps only PSS and the SSS signals to related resource elements. The rest of the packet is filled with zeros. The process can be shown as
$$
a^{j_s}_{k_s,l_s}\sim \mathcal{CN}(0,\sigma_s^2),
$$
where $k_s$ and $l_s$ are also given above. Remaining resource blocks of the jammer are filled with zeros: $a^{j_s}_{k_{rf},l_{rf}}=0$ and $a^{j_s}_{k_m,l_m}=0$. In this case, discrete time representation of the transmitted SSJ signal can be denoted as

\begin{equation}
j_{s}(l)=\frac{1}{\sqrt{K}}\sum_{k=0}^{K-1}{a^{j_s}_{k,l}e^{j2\pi k l/K} \omega\left( {l}/{K+K_{CP}-1} \right)}.
\end{equation}
Note that, the SSJ signal transmission should be synchronized with the eNodeB, and the indices, $k_s$ and  $l_s$, should also be known by the attacker.

\subsubsection{Reference Signal Jamming (RSJ)}
In reference signal jamming (RSJ), the attacker is assumed to know the locations of the reference signals. For RSJ signal generation, we have followed an approach similar to SSJ signal generation. We assumed that the attacker generates the random signal only on the reference signals locations and maps zeros on the data locations similar to the SC. The mapping can be shown as $$a^{j_{rf}}_{k_{rf},l_{rf}}\sim \mathcal{CN}(0,\sigma_r^2), $$ where $k_{rf}$ and $l_{rf}$ are also given above. The remaining resource elements are filled with zeros; $a^{j_{rf}}_{k_s,l_s}=0$ and $a^{j_{rf}}_{k_m,l_m}=0$. In this case, discrete time representation of the RSJ signal is
\begin{equation}
j_{rf}(l)=\frac{1}{\sqrt{K}}\sum_{k=0}^{K-1}{a^{j_{rf}}_{k,l}e^{j2\pi k l/K} \omega\left( {l}/{K+K_{CP}-1} \right)}.
\end{equation}

\subsubsection{Received Signal Model}
In an ideal case, received signal is affected by the channel attenuation and the noise. Considering a flat fading discrete time uncorrelated channel in the existence of a jammer, we can obtain the received baseband signal as
\begin{equation}
r(l)=\frac{s(l)\cdot h(l)}{\sqrt{d_1^{PL}}}+\frac{j_i(l)\cdot g(l)}{\sqrt{d_2^{PL}}}+\eta(l),
\end{equation}
where $j_i(l)$ is the jamming signal of the jammer type $i$, while $i\in\{b,s,rf\}$ respectively for BJ, SSJ and RSJ. $r(l)$ is the received signal, $h(l)$ is the channel coefficient for the transmitted signal, $g(l)$ is the channel coefficient for the jammer signal, $d_1$ is the distance between the eNB and the MN, $d_2$ is the distance between the jammer and the MN, $PL$ is the path loss exponent and $\eta(l)$ is the noise at the receiver. The channel coefficients are assumed to be  complex valued random variables with $h(l),g(l)\sim \mathcal{CN}(0,\sigma^2)$ distribution, and the noise at the receiver is assumed as $\eta(l)\sim \mathcal{CN}(0,\sigma_n^2)$. Resulting signal-to-noise (SNR) ratio expression can be obtained as the following
\begin{equation}
\text{SNR}=\frac{P_s \sigma^2}{\sigma_n^2 d_1^{PL}},
\end{equation}
where $P_s$ denotes the power of the transmitted signal. Another important parameter for jammer analysis is the signal-to-jamming ratio (SJR) and given as the following
\begin{equation}
\text{SJR}=\frac{P_s}{P_{j_i}}{\left(\frac{d_2}{d_1}\right)}^{PL},
\end{equation}
where $P_{j_i}$ is the jamming power.
In the following section, jammer signal identification model will be presented. We aim to identify the jammer type, $i$, by applying different transformation and classification methods to the $r(l)$.

\section{Identification System Model}
The general system structure and alternative operations for the system model are presented in this section. Identification system model is shown in Figure \ref{fig:cell}. The pre-processing part of the identification system is carried out via a TFT which represents a 1D time signal into 2D time-frequency plane. After transformation, 2D output signal can be named as time-frequency representation (TFR). Before classification, the TFR of the received signal is saved as an image. The second part of the identification system includes the automatic classification for the generated images.
\subsection{Pre-processing with Time-Frequency Transformations}

Time-frequency transformations (TFT) allow analyzing the temporal changes of the signal, and they give a compact representation on the time-frequency domain. In his famous overview paper \cite{cohen}, Cohen explains this advantage with an example on sunlight analysis. If the collected data is Fourier transformed, the power density spectrum does not show that the spectral composition of the signal is very different in sunset or sunrise in comparison with other time periods. Therefore, one can suggest that we can capture measurements in predefined time intervals and Fourier transform these samples. By extending or narrowing the time period we can find temporary changes in our data. This example forms the fundamental idea behind the STFT or its magnitude square representation spectrogram, and it also indicates the trade-off between time and frequency resolution. Other Fourier based TFT's can be considered improvements on the spectrogram to provide effective solutions to tackle this trade-off.
\begin{figure*}[t!]
	\centering
	\includegraphics[width=\linewidth, height=0.25\textheight]{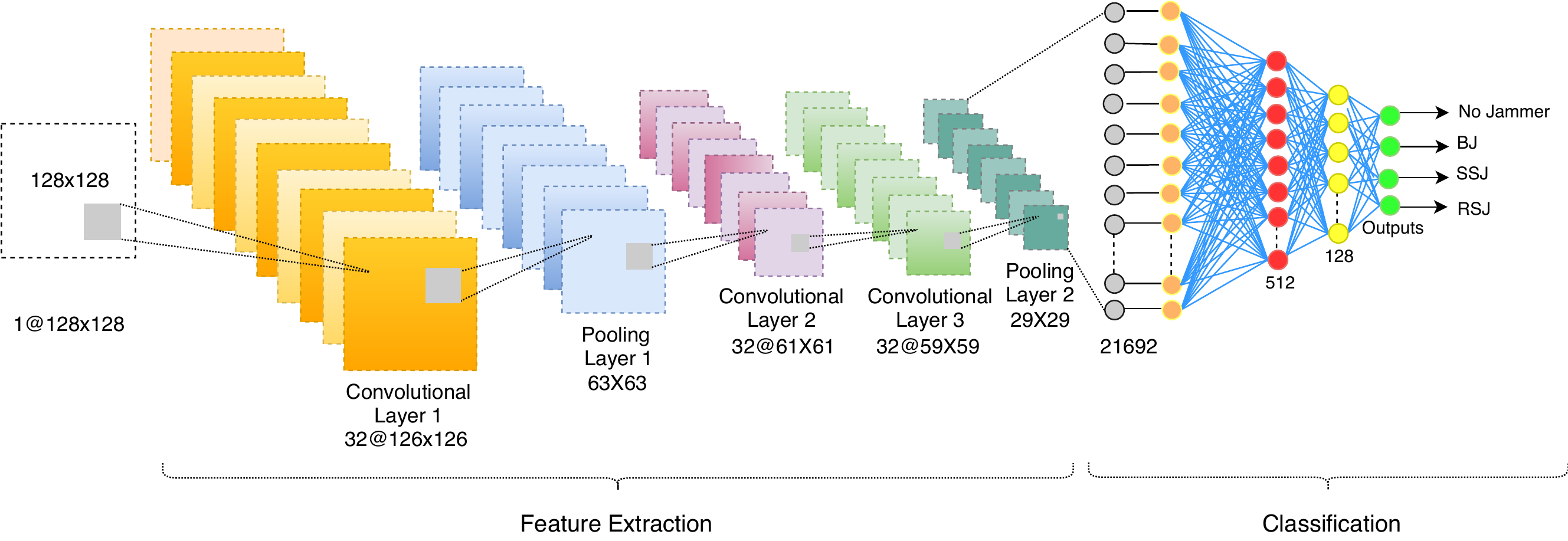}
	\caption{A detailed DCNN block diagram. $F@\Upsilon_1\times\Upsilon_2$ where $F$ represents the number of feature maps in the corresponding layer, whereas $\Upsilon_1\times\Upsilon_2$ denotes the size of anindividual feature map.}
	\label{block}
\end{figure*} 
\subsubsection{Short Time Fourier Transformation (STFT)}
In spectrogram, first a windowing function $\omega(l)$ is applied to the received signal to divide it into short time periods. Then fast Fourier transform (FFT) is applied separately to each time period. We can show the STFT  of the $r(l)$ as

\begin{equation}
R_{s}(c,\kappa)= \frac{1}{\sqrt{2\pi}}\sum_{l=(-L+1)/2}^{(L-1)/2}r(l)\omega(c-l)e^{-j2\pi c l/L},
\end{equation}
where $c=t\times f_s$ indicates the sample number, and $\kappa$ is the frequency variable. For our case, frequency values of the related subcarriers as $\kappa=k\Delta_f$ where $\Delta_f$ shows the frequency difference of the consecutive subcarriers. $\omega(l')$ is a discrete time rectangular function with length $ \tilde{l} $ having a unit magnitude for $(-\tilde{l}+1)/2\leq l'\leq (\tilde{l}-1)/2 $ . Note that $\omega(l')$ can also be generated as a Hamming or Hanning function, but we only consider the rectangular form because of its simplicity. Then the spectrogram of the signal is

\begin{equation}
P_{s}(c,\kappa)=|R_{s}(c,\kappa)|^2.
\end{equation}
Time resolution of the spectrogram is $\delta_{t_S}=(\tilde{l}-1)/(2f_s)$ \cite{figueiredo2004time}. Using this approach, we can obtain the frequency content of the signal for short time periods.

\subsubsection{Choi-Williams Transformation (CWT)}
First proposed in \cite{choi}, Choi-Williams Transform (CWT) actually aims to overcome difficulties on another TFT named as Wigner-Ville transformation (WVT). In WVT, multitone signals generate high power cross terms that should be zero. CWT remarkably  reduces the cross-terms without worsening the spectral representation. The CWT of the received signal is
\begin{equation}
R_{cw}(c,\kappa)=2\sum_{l=0}^{2L-1}S'(c,l)e^{-j2\pi \kappa l/L},
\end{equation}
where $S'(c,l)$ is
\begin{equation}
\small
S'(c,l)=
\begin{cases}
S(c,l) \text{ , }& 0\leq l \leq L-1 \\
0 \text{ , } &l=L \\
S(c,l-2L) \text{ , } &L+1\leq l \leq 2L-1
\end{cases}
\end{equation}
and
\begin{equation}
S(c,l)= \omega(\nu)\sum_{\nu=-L/2}^{L/2}\frac{1}{\sqrt{4 \pi l^2/ \sigma}}e^{-\frac{(\nu-1)^2}{4 \pi ^2/ \sigma}}r(l+\nu)r^*(l-\nu),
\end{equation}
Similar to spectrogram, we can define the absolute square of the transform:

\begin{equation}
P_{cw}(c,\kappa)=|R_{cw}(c,\kappa)|^2.
\end{equation}
The effective time resolution is $\delta_{t_{cw}}=(\tilde{l}-1)/(2f_s)$ \cite{figueiredo2004time}. Finally the resulting TFR $P_{cw}(c,\kappa)$ is color-mapped and saved for the further classification process.

\subsubsection{Gabor Wavelet Transformation}
In the presence of a smart jammer, the received signal $r(l)$ would show varying characteristics over time. While the attacked parts exhibit abrupt changes in the signal, non-attacked parts would have smooth changing components. These abrupt changes become apparent in higher frequency components. Increasing the scaling factor of the wavelet provides a better representation of the abrupt changes on the signal. We can construct a TFR from the received signal by sequencing the arrays obtained from different scale factors of wavelet transformations.

Among  various  wavelet  bases,  Gabor  functions provide  the  optimal  resolution  in  both time  and  frequency  domains by using a Gaussian shaped windowing function \cite{cohen}. Gabor wavelet transformation function can be expressed as
\begin{equation}
\Psi(c,f)=\frac{f}{\gamma \sqrt{\pi}}e^{\frac{f^2}{\gamma^2 }c^2}e^{j2 \pi cf},
\end{equation}
where $f/\gamma$ is the scaling factor and $\gamma$ is the quality factor and $f=1,2,\cdots,F$. The transformation is carried out via convolving the received signal $r(l)$ with defined Gabor wavelet transformation function:
\begin{equation}
R_{gm}(c,f)=r(l)*\Psi(c,f),
\end{equation}
where $*$ denotes the convolution operation and  $f$ shows the different frequency values on which wavelet transform is applied. The TFR, also named as scalogram for Gabor-wavelet transform is

\begin{equation}
P_{gm}(c,f)=|R_{gm}(c,f)|^2,
\end{equation}
where the time resolution of the scalogram is defined as $\delta_{t_{gm}}(f)=\gamma/f$ \cite{figueiredo2004time}. $P_{s}$, $P_{cw}$ and $P_{gm}$ are TFR's of the received signal and they are constructed as 2D matrices. They are stored in the memory for training the classification methods. The values in these magnitude level representations are color-mapped by their amplitude values. The color-mapping operation can be seen as a quantization of the magnitude levels into different color levels.

\subsection{Classification Methods}
In this work, we consider SVM and DCNN for classification purposes due to their high performances as reported in the literature \cite{S8}, \cite{wu2017jamming}.
\subsubsection{Support Vector Machine (SVM)}
Support vector machine (SVM) is a supervised learning method and  was firstly developed for binary classification. The main idea behind SVM is finding the optimal decision boundary  by discriminating the feature vectors. The decision boundary is also named as a hyperplane and for optimality it has to maximize the separation between two data classes. The method starts with selecting a hyperplane in the feature space and is defined as follows;

\begin{equation}
\textup{\textbf{w}}^{T}\textup{\textbf{x}}+\,b=0 ,
\end{equation}
where $\textup{\textbf{x}}$ is the feature vector or in this study output of PCA, $\textup{\textbf{w}}$ is the support vector, ${({\cdot})}^{T}$ denotes transpose operation and $b$ is the bias term. SVM separates the data classes by maximizing the margin defined as minimum distance of any data points to the decision boundary. If the data are not linearly separable, optimization problem is defined as below to find  the maximal margin hyperplane;
\begin{equation}
\begin{matrix}
&\mkern-18mu\mkern-18mu\min\limits_{\xi,\textbf{w},b}  \quad  \frac{1}{2}  \|\textup{\textbf{w}}\|^{2}+ \lambda \sum_{p=1}^{m}\xi_{p}\\
&\text{subject to}  \;\;\; y^{(p)}(\textup{\textbf{w}}^{T}$\textup{\textbf{x}}$^{(p)}+\,b) \geq 1-\,\xi_{p}\bigskip\\
&	\quad\quad\quad\quad\quad\;\; \text{and} \;\; \xi_{p} \geq 0   \; \; p=1,...,m.
\end{matrix}
\end{equation}
where $m$ is the number of training samples,  $\mathbf{x}^{(p)}$, $\mathbf{y}^{(p)}$,   $\xi_{p}$ are related to $p^{\text{th}}$ training sample and $y^{(p)}$ is a class label, which has one of only two values, either $-1$ or $1$.  $\xi_{p}$ is a slack variable. The first term minimizes the distance to the closest data point, and the second term reduces the number of misclassified points. Optimization problem  is constructed as the Lagrangian as given below

\begin{equation} \label{eq:s1}
\begin{matrix}
\mkern-18mu\mkern-36mu\mkern-36mu\mkern-36mu\mathcal{L}(\textup{\textbf{w}},b,\textup{\textbf{a}})=\frac{1}{2}  \|\textup{\textbf{w}}\|^{2} + \lambda\sum_{p=1}^{m}\xi_{p} \bigskip\\
-{\sum_{p=1}^{m} a_{p} [y^{(p)}(\textup{\textbf{w}}^{T}\textup{\textbf{x}}^{(p)}+\,b) -\,1 + \xi_{p}]}-\,{\sum_{p=1}^{m} \mu_p \xi_{p}},
\end{matrix}
\end{equation}
where $ 0 \leq a_{p} \leq \lambda $, $\lambda$ is a hyperparameter called as penalty of the error term or regularization term, and  $\mu_p \geq 0$  is the Lagrange term. When the above equation is solved, it turns into following form;

\begin{equation} \label{eq:scorefunc}
\begin{matrix}
f(x)={\sum_{p=1}^{m} a_{p}  y^{(p)} \mathcal{K}(\textup{\textbf{x}}^{(p)},\textup{\textbf{x}})+b},
\end{matrix}
\end{equation}
where $\mathcal{K}$ is a kernel function, and $f(x)$ is called as decision function or score function which is used to compute score  for each input vector. According to the output of (\ref{eq:scorefunc}), SVM predicts the class of each input.
\begin{figure*}[hbt]
	\centering
	\includegraphics[width=0.95\linewidth]{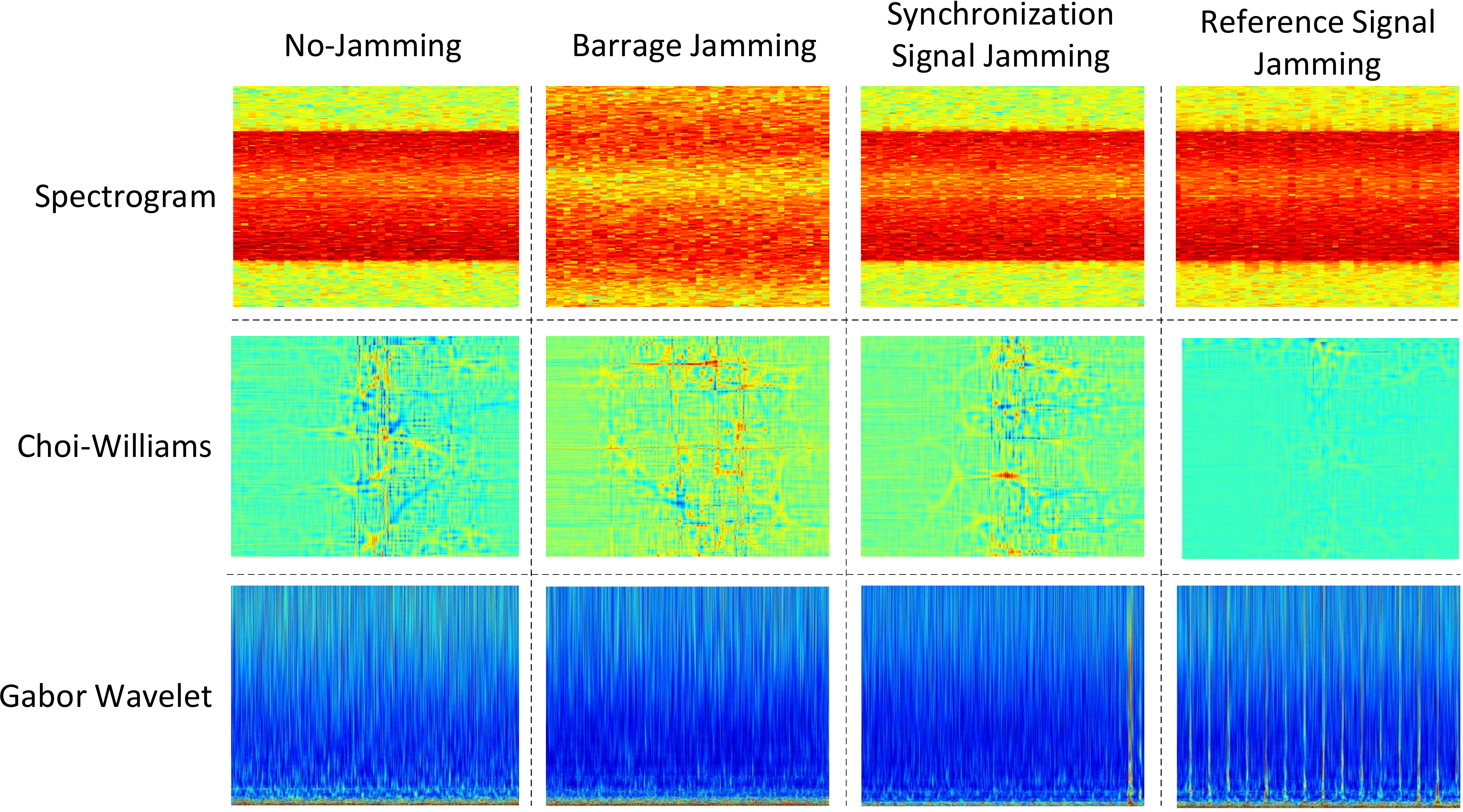}
	\caption{Spectrogram, CWT and Gabor wavelet transform applied images. (from left to right no-jamming, BJ, SSJ and RSJ; from up to down  spectrogram, CWT, Gabor wavelet transform)}
	\label{fig:numerical}
\end{figure*}

\subsubsection{Deep Convolutional Neural Networks (DCNN)}
Deep convolutional neural networks (DCNN) can be thought of as a series of layers which are trained by the network model using a given dataset. These layers are convolutional layers, sub-sampling layers (also known as pooling layers) and full-connected layers. Pooling and convolution layers are fundamental layers of feature extraction part of DCNN. From the first  layer to the last one,  simple features are extracted initially, and gradually it continues to more complex features. The convolution layer is based on a discrete convolution process. Discrete convolution in two dimensions is given as the following;

\begin{equation}
Y(r,t)=\sum_{k_1=-\infty}^{\infty}\sum_{k_2=-\infty}^{\infty}   X(r-k_1,t-k_2)W(k_1,k_2),
\end{equation}
where $X_{n_{1}\times n_{2}}$  represents two dimensional input matrix and $W_{m_{1}\times m_{2}}$  is a kernel/filter matrix with $m_{1}\leq n_{1}$ and $m_{2}\leq n_{2}$ . A feature map is produced by sliding the convolution filter over the input signal. The sliding scale is a hyper-parameter known as stride. When zero padding is not used, calculation of feature map size or convolution layer output length  for each dimensions can be realized using the following equation
\begin{equation}
o_{d} =  \cfrac{n_{d}-\,m_{d}}{s_{d}} +\, 1  \;\;\;\;\; d=1,2.
\end{equation}
$n_{d}$ and $m_{d}$ represent the length of the input vector and the kernel length in $d^{\text{th}}$ dimension, where $s$ is  the value of stride. Pooling operations decrease the size of feature maps with some functions  taking  average or maximum value of each distinct region of size $a_1 \times a_2$ from the input.  Pooling layer solves disadvantages related to probability of over-fitting and computational complexity \cite{S17_book}. Here, pooling layer does not include learnable parameters like bias units or weights.  After the feature extraction, classification part follows and consists of one or more full connected layers.  In full connected layers,  each neuron is connected to the preceding layer, and at the end of the classification neural network produces the outputs.

In neural networks, activation function is used for producing a non-linear decision boundary via functions of the weighted inputs. As often preferred, in this work rectified linear unit activation (ReLu) is  selected and it is defined as follows:
\begin{equation}
\phi(z)= \text{max}(0,z).
\end{equation}
In output layers especially related to the classification problem, softmax function is a very strong option. As a normalized function, the softmax function is useful to obtain meaningful class membership predictions in multiclass settings  \cite{S17_book}. The softmax function is defined below, where $z$ is a vector of the inputs, $i$ is the output index, and  $M$ is the number of classes.

\begin{equation}
p(y=i|z)=\phi(z)=\dfrac{e^{z_{i}}}{\sum_{n=1}^{M}e^{z_{n}}}.
\end{equation}

\section{Numerical Analysis}

LTE signal generation and transmission scheme is modeled with the MATLAB LTE Toolbox. It offers a realistic modeling and testing framework for LTE networks. The parameters regarding to complete system can be found in Table \ref{table:parameters}. The channels between radio nodes are assumed as flat fading with unit gain Rayleigh distribution.

Four different jamming scenarios are considered, and the classification labels are: the absence of a jammer (no jamming), BJ, SSJ, and RSJ. The jamming is assumed to be active throughout a packet transmission duration. The transmission is repeated 5000 times for each scenario. In each scenario, the received signal $r$ is generated in accordance with (6).

\begin{table}[t]
	\footnotesize
	\centering
	\caption{Parameters for numerical analysis.}
	\label{table:parameters}
	\begin{tabular}{|p{1.8cm}|l|l|}
		\hline
		Step                                                              & Parameter              & Value                     \\ \hline
		& \# of subcarriers      & 140                       \\ \cline{2-3}
		& \# of subframes        & 10                        \\ \cline{2-3}
		& Duplex Mode            & FDD                       \\ \cline{2-3}
		& Cyclic Prefix          & Normal                    \\ \cline{2-3}
		& Modulation Type        & QPSK                      \\ \cline{2-3}
		& $d_1$                  & 1, 1.5                    \\ \cline{2-3}
		& $d_2$                  & 1, 1.5                    \\ \cline{2-3}
		& SNR (dB)                   & 0, 5                 \\ \cline{2-3}
		& SJNR (dB)                   & -5, 0, -5                  \\ \cline{2-3}
		& Channel Fading         & Rayleigh                  \\ \cline{2-3}
		& Channel  Gain          & 1                         \\ \cline{2-3}
		& Sampling Frequency          & 1.92 MHz                  \\ \cline{2-3}
		& $PL$                   &4                          \\ \cline{2-3}
		& $n_1$                       &128                           \\ \cline{2-3}
		\multirow{-15}{*}{ \begin{tabular}[c]{@{}c@{}} Signal\\Transmission\end{tabular} }    & $n_2$                       &128                           \\ \hline \hline
		& $\sigma_{cw}$          & 100                       \\ \cline{2-3}
		& $\gamma$               & $5/2\sqrt{2 \text{log}2}$ \\ \cline{2-3}
		& F                      & 256                       \\ \cline{2-3}
		& $\delta_{t_s}$ , $\delta_{t_{cw}}$          & 0.1024                       \\ \cline{2-3}
		& $\delta_{t_{gm}}$           & $3.221/f$                     \\ \cline{2-3}
		& L                     & 512                       \\ \cline{2-3}
		\multirow{-5}{*}{ \begin{tabular}[c]{@{}c@{}} Time-Frequency\\Transforms\end{tabular}  } &   Colormap                     &   Jet                          \\ \hline \hline
		& PCA component number   & 1000                      \\ \cline{2-3}
		& C                      & 1                          \\ \cline{2-3}
		& Kernel function        & linear                    \\ \cline{2-3}
		& Gamma value            & 0.001                     \\ \cline{2-3}
		\multirow{-5}{*}{\begin{tabular}[c]{@{}c@{}} PCA-SVM\\Classification\end{tabular}   }  & Batch size             & 64                         \\ \hline
		& Epoch number           & 40                         \\ \cline{2-3}
		& Convolution            &  \tabularnewline
		& layer stride           & (1,1)                     \\ \cline{2-3}
		&                        & Stochastic \tabularnewline                                             & Optimizer              & Gradient Descent            \\ \cline{2-3}
		& Learning rate          & 0.1                        \\ \cline{2-3}
		&                        & Categorical cross \tabularnewline
		& Loss Function                       & entropy function            \\ \cline{2-3}
		& Total parameters           & 13,867,988                  \\ \cline{2-3}
		& Trainable parameters       & 13,866,508                  \\ \cline{2-3}
		\multirow{-4}{*}{ \begin{tabular}[c]{@{}c@{}} DCNN\\Classification\end{tabular}}     & Non-trainable params   &1,480                    \\ \hline
	\end{tabular}
\end{table}

Same process is repeated considering four different combinations of the jammer and the MN locations. The location cases and their equivalent SNR/SJR values with error vector magnitude (EVM) values are given in Table \ref{table:index}. As indicated in \cite{Husey}, EVM is an error metric that strongly depends on the signal-to-noise/interference ratio (SINR). $C_1$ corresponds to the case when the jamming signal does not significantly reduce the transmission. On other cases, EVM values are higher than $70\%$, indicating that the transmission is blocked by the jamming signal. The transmitted signal is assumed to have unit power. SNR/SJR values are obtained by changing the variance of the noise signal and jamming power.  For all location cases,  radio nodes are assumed to be stationary.

\begin{table}[t]
	\footnotesize
	\caption{Description of location cases and their distances, SNR, SJR and SJNR equivalents}
	\label{table:index}
	\centering
	
	\begin{tabular}{l|l|l|l|l|l|l|l|l|}
		\cline{2-9}
		& \multirow{2}{*}{$d_1$} & \multirow{2}{*}{$d_2$} & \multirow{2}{*}{SNR} & \multirow{2}{*}{SJR} & \multicolumn{4}{c|}{EVM (\%)} \\ \cline{6-9} 
		&  &  &  &  & No & BJ & RSJ & SSJ \\ \hline
		\multicolumn{1}{|l|}{$C_1$} & 0.5 & 1.5 & 10 & 10 & 16.58  & 18.25 & 17.69 & 17.14 \\ \hline
		\multicolumn{1}{|l|}{$C_2$} & 1 & 1 & 5  & 0 & 23.14  & 95.82 & 85.18 & 77.64 \\ \hline
		\multicolumn{1}{|l|}{$C_3$} & 1 & 1.5 & 5 & 5 & 23.14 & 86.15  &  69.22 & 71.92 \\ \hline
		\multicolumn{1}{|l|}{$C_4$} & 1.5 & 1 & 0 &-5  & 30.58 & 99.72  & 92.19 & 79.14 \\ \hline
		\multicolumn{1}{|l|}{$C_5$} & 1.5 & 1.5 & 0 & 0 & 30.58  & 89.18 & 83.26 & 75.84 \\ \hline
		\multicolumn{1}{|l|}{$C_6$} & 1.5 & 0.5 & 0 & -10 & 30.58 & 99.16 & 96.24  & 98.47 \\ \hline
	\end{tabular} 
\end{table}

At the second step of the analysis, three different TFT's are applied on the received signals in order to compare their performances with classifiers. 
Figure \ref{fig:numerical} shows the images obtained from the $P_{s}$, $P_{cw}$ and $P_{g}$ on logarithmic scale. These images are saved in portable network graphics (png) format and fed into two classifiers. All images for a given transformation are cropped to the same size in order to eliminate misleading features.

As it can be seen from Figure \ref{fig:numerical}, the existence of BJ is observable for every pre-processing method. Considering spectrogram, differences among no-jamming, SSJ and RSJ are not apparent due to the single resolution property as discussed in Section III. Since most of the energy is located near the low frequencies, the abrupt changes occuring in high frequencies are not observable from the generated images. For CWT, the difference between no-jamming and BJ are not as apparent as spectrogram and Gabor wavelet transform, because of the obscure background. As discussed above,  the ambiguity arises from the cross-term effects is still observable, despite the cross-term minimizing effects of CWT. Yet, classification results in the following of this paper show that the features in the images are still separable by the classifiers.

In Gabor wavelet transform the images of four different jamming scenarios are separable from each other. For the no-jamming scenario, energy is mostly localized in the low-frequency values. As it can be seen from the turquoise colored region, received signal contains abrupt changes due to the channel fading and the noise. These changes are closely located energy content on the time-frequency domain. Hence, these changes are represented with the same color. In BJ scenario, jammer attacks are spread over the entire time-frequency plane. As the power of the jammer increases, the abrupt changes on the signal become dominant over the time-frequency plane. In SSJ and RSJ, the attacked locations on time-frequency plane become apparent due to the multi-resolution property of wavelet transform. The emitted energy from SSJ and RSJ becomes apparent in Gabor wavelet transform since the wavelets are localized on time-frequency plane.

In classification part, we compare two machine learning algorithms; DCNN and SVM. DCNN architecture is implemented in KERAS Python library interface that works with Tensorflow back end, while SVM is implemented in Phyton.  We train each model with 1000 images and the size of images are $128\times128$. On parameter selection, we utilize the accuracy of classification as the main metric. As shown in Figure \ref{block}, DCNN architecture consists of three convolution layers, two pooling layers and three full-connected layers. We choose $2\times2$ for pooling size and $3\times3$ filter size  for convolutional kernels. After each layer we apply batch normalization and ReLu function, except the output layer in which softmax function has been used. Also we employ dropout layer of $ 40\% $ rate  to prevent over-fitting. On the training of the model we use cross entropy $J(\varTheta)$ as a cost function  \cite{S18_book}:
\begin{figure}[tb]
	\centering
	\includegraphics[width=\linewidth]{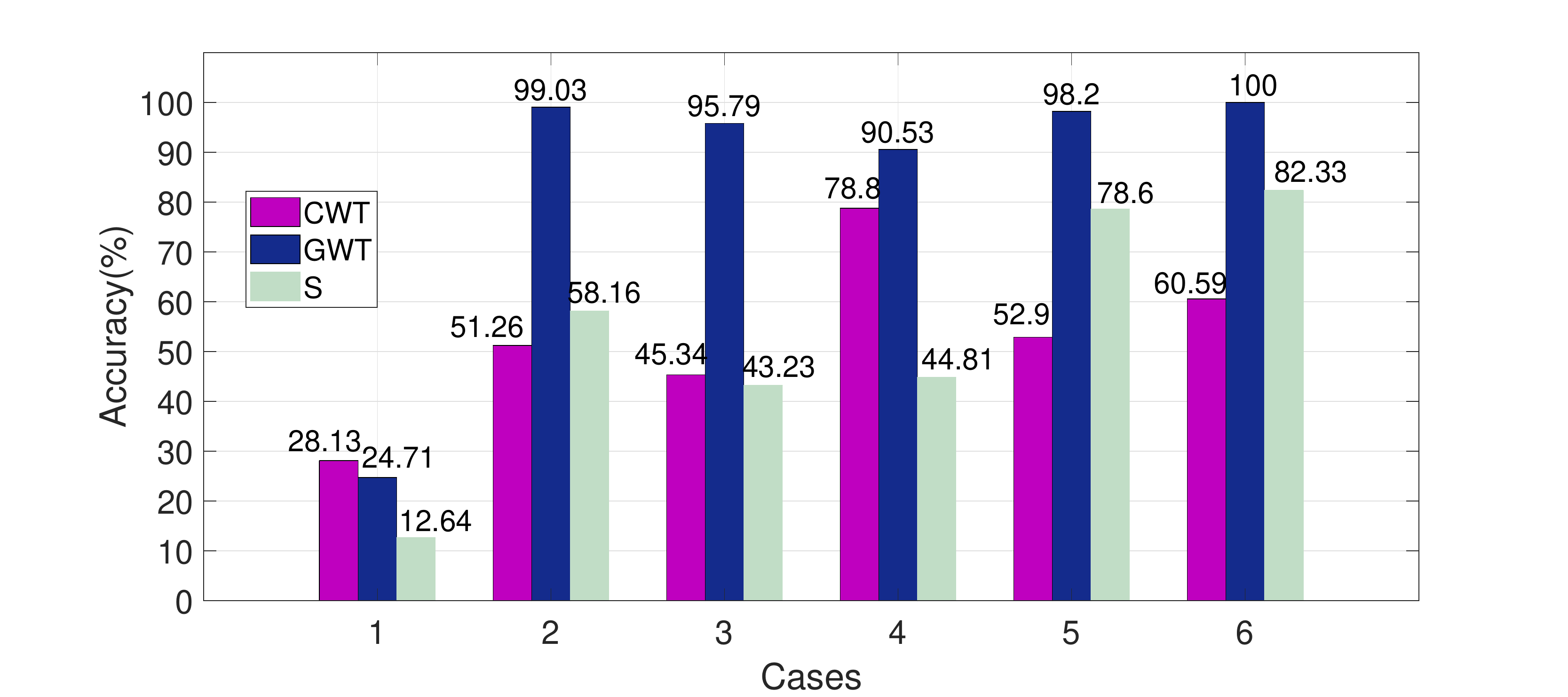}
	\caption{ Different time-frequency analysis comparison for SVM classification.}
	\label{fig:svm}
\end{figure}

\begin{figure}[tb]
	\centering
	\includegraphics[width=\linewidth]{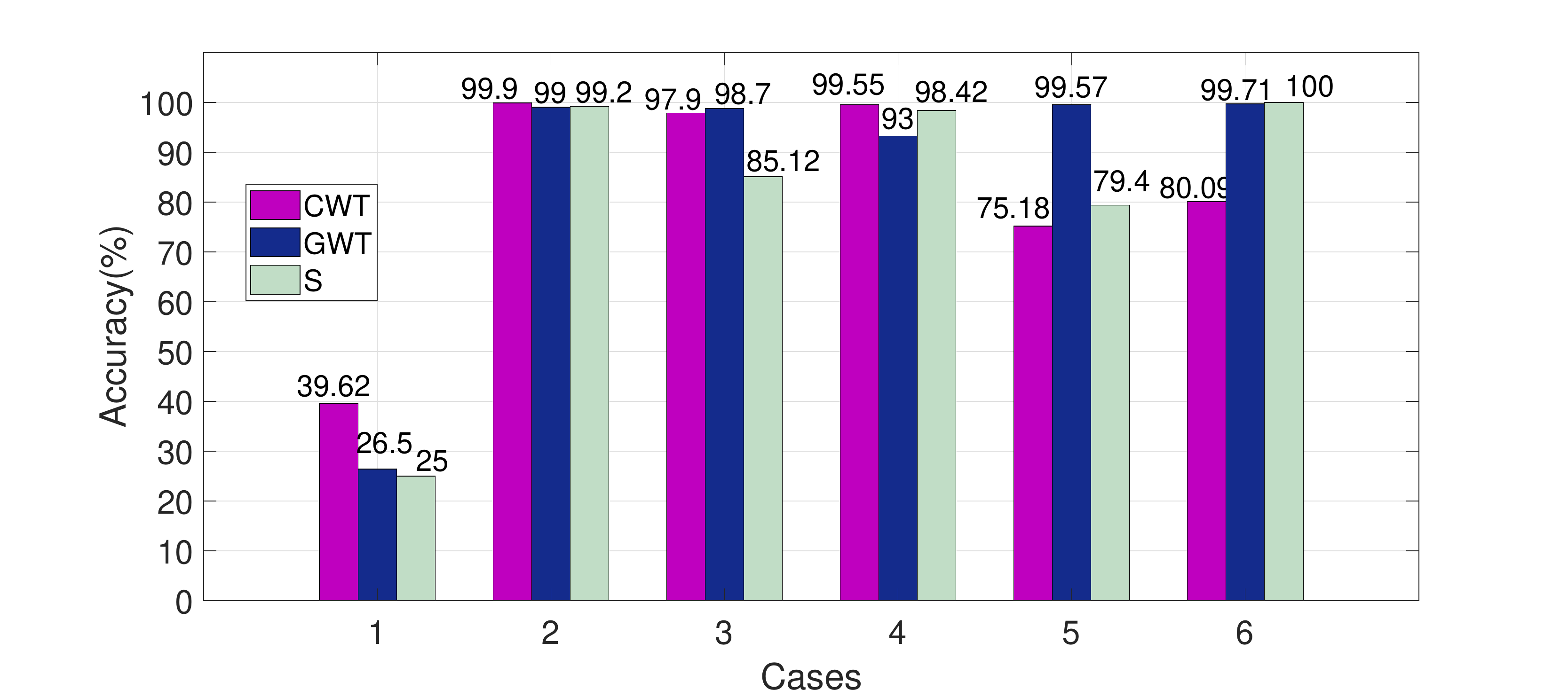}
	\caption{Different time-frequency analysis comparison for DCNN classification.}
	\label{fig:dcnn}
\end{figure}

\begin{figure*}[t]
	\begin{center}	
		\subfigure[]{
			\label{confusion-P1}
			\includegraphics[width=0.6\columnwidth]{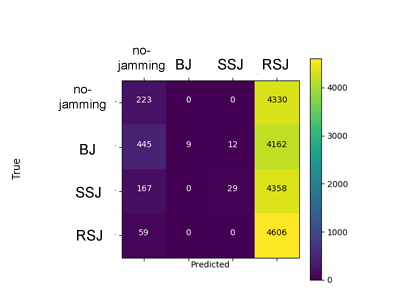} }  
		\subfigure[]{
			\label{confusion-P2}
			\includegraphics[width=0.6\columnwidth]{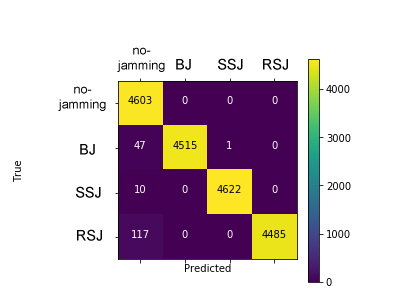} } 
		\subfigure[]{
			\label{confusion-P3}
			\includegraphics[width=0.6\columnwidth]{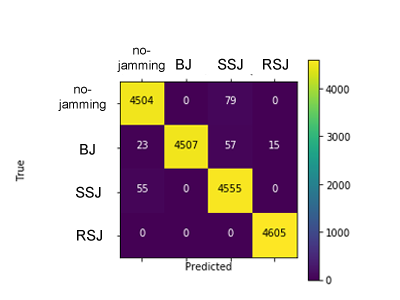} } \\
		\subfigure[]{
			\label{confusion-P4}
			\includegraphics[width=0.6\columnwidth]{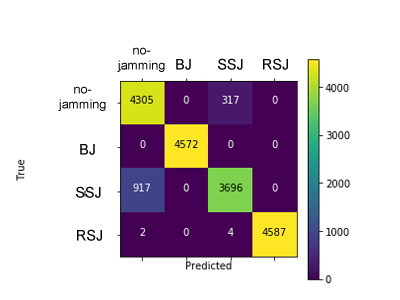} }  
		\subfigure[]{
			\label{confusion-P5}
			\includegraphics[width=0.6\columnwidth]{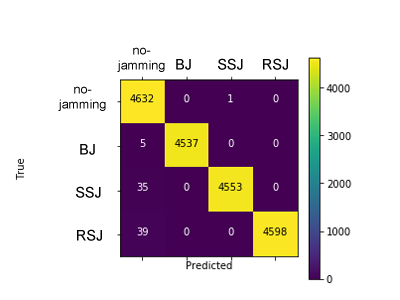} } 
		\subfigure[]{
			\label{confusion-P6}
			\includegraphics[width=0.6\columnwidth]{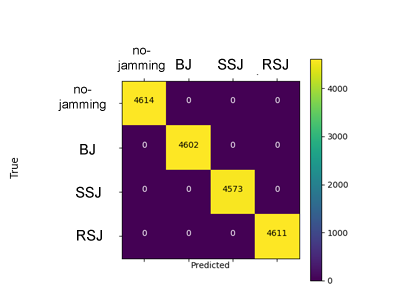} } 
	\end{center}
	\vspace{-0.4cm}
	\caption{Confusion matrices when Gabor wavelet transform is applied before DCNN classifier for respectively (a) $C_1$, (b) $C_2$, (c) $C_3$, (d) $C_4$, (e)  $C_5$ and (f)  $C_6$}
	\label{confusion}
\end{figure*}

\begin{equation} \label{eq:cross_entropy}
J(\varTheta)= \frac{-1}{m} {\sum_{p=1}^{m} \sum_{n=1}^{M} y_n^{(p)} \text{log}(q_n^{(p)})}.
\end{equation}
where $M$ is the number of classes and $m$ is number of instances. $q_n^{(p)}$ is the estimated probability that the instance $x^{(p)}$ belongs to class $k$. $y_n^{(p)}$ is equal to 1 if the target class for the $p^{\text{th}}$ instance is $k$. Otherwise, it is equal to 0. In the next step, we compute the gradient vector for every class and  use stochastic gradient descent optimizer to find the parameter matrix $\varTheta$ which minimizes the cost function.
Besides DCNN, we also applied SVM to compare their classification accuracies. Unlike DCNN, SVM requires an explicit feature extraction step to realize classification. Principal component analysis (PCA) is used as an effective and robust method for feature extraction  \cite{Pal}. The central idea of PCA is to reduce the dimensionality of the data set in which there are a large number of interrelated variables, while retaining as much as possible of the variation present in the data set \cite{jobook}. After feature extraction through PCA, the dimension of the input vector $\mathbf{x}$ fed to the classifier is equal to the number of extracted principle components. SVM is applied to the multi-class classification problem by using the "one against the rest" approach [53]. The fundamental idea underlying this approach is to realize the multi-class classification by using multiple binary SVMs collectively. After completion of the classification model training, 18400 test images are used to determine the performance of the algorithms for  identifying the jammer types considering a single position case. For every position case, same training and test process is repeated. 
Classification accuracy results obtained by SVM and DCNN for different simulation setups can be seen from Figure \ref{fig:svm} and Figure \ref{fig:dcnn}.

Figure \ref{fig:svm} shows the classification accuracies when SVM is selected as the classifier. The groups at horizontal axis show the case numbers, where their SNR/SJR equivalents can be found from Table \ref{table:index}. Considering $C_1$, the identification accuracies are lower than other cases. In this case, jamming signal is not detected by the identification system. However note that, the jamming signal cannot block the transmission anyways, which makes identification obsolete at the first place as the identification accuracies are lower than other cases. Considering other cases, the figure shows that  the Gabor wavelet transform outperforms other classification methods with a minimum of 90\% accuracy. As the signal and the jamming powers become closer to each other, in other words as the SJR approaches to zero, the classification accuracy of the Gabor wavelet transform increases. In high SNR values, classification accuracy is higher than the low SNR values considering Gabor wavelet transform. Considering $C_6$, the power of the jamming signal is much more powerful than the message signal. In this case, jammer signal becomes dominant over the time-frequency plane, where identification accuracy improves for all pre-processing methods.

Interestingly, other transform methods do not follow the same accuracy behavior as the Gabor wavelet transform. Both Choi-Williams transform and spectrogram work better in the low SNR region. Especially Choi-Williams works best when SJR is low. This is plausible because when the power of the jamming signal is higher than the message signal, the jamming patterns become more apparent.

Figure \ref{fig:dcnn} shows the classification accuracies when DCNN is selected as the classifier. Similar to the results of SVM classifier, the jammer activity cannot detected by the DCNN classifier in $C_1$. In other cases, all transformation methods perform better than the SVM classifier in this case. Gabor wavelet transform gives  the best accuracy results overall with a minimum 93\% accuracy. Similar to the SVM classification, Gabor wavelet works better as the SJR approaches zero.  Although the remaining transformation techniques show high performance for $C_2$ and $C_4$, their performance are steadily reduced in $C_3$ and $C_5$.

If we evaluate the overall success of the classification methods, the first noticeable feature is the superior  jammer classification performance of the DCNN for all different transforms and setups. 
Although SVM has respectable performance in certain cases, DCNN with its hidden layers provides consistent and robust performance for all differing setup choices and TFT methods.

Considering the results in Figure \ref{fig:svm} and \ref{fig:dcnn} together, best working identification method would be the combination of {Gabor wavelet} transform and DCNN classifier. {As expected from the previous sections}, Gabor wavelet transform gives the best representation among other transformation methods because the wavelet characteristics alter the time-frequency plane resolution and highlight abruptions resulting from jamming.

Figure \ref{confusion} shows the confusion matrices of Gabor wavelet transform when used with the DCNN classifier. These confusion matrices describe the performance of the classification model by demonstrating the comparison of model prediction and true class for the test dataset. It is a quality measure for the class by class identification performance of the proposed method. In $C_1$, DCNN can not accurately differentiate between the classes, because the very low powered jamming signal can not get detected on the time-frequency plane. In $C_2$, the most commonly confused classes are RSJ and no-jammer. One important observation from all of the cases is that there is almost no confusion between the different jammer types. 

\section{Open Issues}

While the proposed system model determines the jammer type and its existence, the anti-jamming strategies can be proposed considering the identification model. Especially, legitimate nodes can obtain an advantage through game theoretical models.

This system model can also be applied for the timing behaviors of the jammers. In this case, attacks can be identified  at MAC layer. As these jamming attacks strictly rely on the time characteristics of the channel, the correlation of consecutive channel taps should be considered for the identification performance. Especially for the hybrid jammer identification, correlated channel models would be a better representation for modeling the real-time attacks.

The proposed identification system can also be extended to multi-antenna systems. The monitoring node can benefit from the diversity came within the multiple antenna structure. The observations from multiple antennas can be fused with different signal processing techniques for jammer identification and localization.

\section{Conclusion}
In this paper, we present a novel jammer identification system and determine the identification performance  over various jamming attack cases. The proposed system can differentiate three main types of jamming attacks: barrage jamming, synchronization signal jamming and reference signal jamming cases along with the absence of jammer case. Even though barrage jammer can be easily detected as an unexpected random noise combined with the original signal, other types of attacks are harder to detect {and to tolerate} since they effectively hide in the targeting time-frequency plane and possibly disrupt communication with a low power. The proposed system model is composed of a wavelet-based pre-processing step and a deep learning based classification stage. We consider an LTE downlink communication scenario, where the effectiveness of the wavelet transform based approach is clearly observed, even in the presence of smart jamming attacks. Considering different pre-processing methods, the superior performance of DCNN is observed in comparison to SVM.

\bibliographystyle{IEEEtran}
\bibliography{ref}

\end{document}